\documentclass[aps,prb,twocolumn]
{revtex4}
\usepackage{graphicx}
\usepackage{amsmath}
\usepackage{xcolor}

\def\be{\begin{equation}}       \def\ee{\end{equation}}
\def\bea{\begin{eqnarray}}      \def\eea{\end{eqnarray}}
\def\ba{\begin{array} }
\def\ea{\end{array} }
\def\bnum{\begin{enumerate} }
\def\enum{\end{enumerate}}

\def\=>{\Rightarrow}
\def\>{\rightarrow}

\def\eye2{Fathbb{I}}

\def\d0{\Delta_{0}}

\begin{document}
\title{\bf Generalization of Anderson's Theorem for Disordered Superconductors}
\author{John F. Dodaro and Steven A. Kivelson}
\affiliation{Department of Physics, Stanford University, Stanford, CA 94305-4060, USA}

\begin{abstract}
We show that at the level of BCS mean-field theory, the superconducting $T_c$ is always increased in the presence of disorder, regardless of order parameter symmetry, disorder strength, and spatial dimension.  This result reflects the physics of rare events -- formally analogous to the problem of Lifshitz tails in disordered semiconductors -- and arises from considerations of spatially inhomogeneous solutions of the gap equation.  So long as the clean-limit superconducting coherence length, $\xi_0$, is large compared to disorder correlation length, $a$, when fluctuations about mean-field theory are considered, the effects of such rare events are small (typically exponentially in $[\xi_0/a]^d$);  however, when this ratio is $\sim 1$, these considerations are important.  The linearized gap equation is solved numerically for various disorder ensembles to illustrate this general principle.
\end{abstract}

\maketitle


Most theoretical analyses of the effect of disorder on the superconducting transition rest on two separate assumptions:  the pairing instability is treated in the context of a generalized BCS mean-field theory, and the disorder is treated either perturbatively or in an effective medium approximation.  The primary focus of this paper is the demonstration that there is additional universal structure to the mean-field solution of this problem when the effect of disorder is treated exactly.  In particular, the presence of rare regions -- which are neglected in effective medium approximations -- leads to the unintuitive conclusion that disorder always increases the mean-field $T_c$, independent of the nature of the disorder and whether we are considering a conventional (s-wave) or unconventional (p-wave or d-wave) superconductor.  We demonstrate this explicitly with numerical solutions of the linearized gap equation for various disorder ensembles.

There is a separate issue of whether this improved analysis of the BCS equations is physically significant.  When the regions that support a local superconducting order parameter at mean-field level are sufficiently rare, the inclusion of fluctuation effects beyond mean-field theory reveals the inferred high transition temperatures to be artifacts of mean-field theory;   in this case, an incorrect solution of the mean-field equations can give a more physically sensible solution.  However, when the $T=0$ superconducting correlation length, $\xi_0$, is comparable to the disorder correlation length, $a$, locally superconducting solutions of the mean-field equations can lead to significant consequences -- such systems exhibit a broad fluctuational regime in which a form of self-organized granularity arises.  We stress this is unavoidable in superconductors with correlation lengths comparable to the lattice constant, even if the disorder is entirely homogeneous as in a ``perfect'' substitutional alloy.


\section{The model}   \label{sec:model}
We consider the tight-binding model
\be
H=-\sum_{a,b} t_{ab} c_a^\dagger c_b + \sum_{ab b^\prime a^\prime}V_{ab b^\prime a^\prime} c_a^\dagger c_b^\dagger c_{ b^\prime} c_{a^\prime}
\ee
where $a=(\vec r,\sigma)$ is an index specifying the lattice position, $\vec r$, the spin polarization, $\sigma$, and possibly -- if relevant -- an orbital index, and $c^\dagger_a$ creates an electron in single-particle state $a$.   We assume that both $t$ and $V$ are short-range in space, but they need not be translationally invariant.  Thus, to fully specify the problem we need to specify the ensemble which determines the probability of different ``configurations,'' {\it i.e.} realizations of $H$.  We will imagine that the ensemble has statistical translational symmetry, so that the configuration-averaged version
\be
{\overline H} = - \sum_{\vec r,\vec r^\prime,\sigma}{\overline t(\vec r-\vec r^{\ \prime})} c_{\vec r,\sigma}^\dagger c_{\vec r,\sigma} 
+ \sum_{\vec r,\vec r^\prime }{\overline V(\vec r -\vec r^{\ \prime})} \hat n(\vec r) \hat n(\vec r^{\ \prime})
\ee
is translationally invariant.  (For simplicity, we have ignored spin-orbit coupling and assumed a density-density interaction, $\hat n(\vec r) = \sum_\sigma c^\dagger_{\vec r\sigma}c_{\vec r\sigma}$, although the same considerations apply to more general models.)  Then when we ask about the effect of disorder on $T_c$, we are comparing the case in which we replace $H$ by ${\overline H}$ (translation symmetry restored) with the full problem using the configuration-dependent $H$.  We will always focus the analysis on systems in the thermodynamic limit, and will assume that the disorder ensemble is sufficiently well-behaved that the system properties are self-averaging in this limit.  Since in general $V$ includes effective interactions obtained by integrating out high energy degrees of freedom, we consider disorder in $V$ as well as single-particle disorder $t_{aa}$.  (See, for example, Ref. \onlinecite{nunner_2005, andersen_2006})

\subsubsection*{BCS mean-field theory} The interacting problem can be treated at mean-field level by introducing the trial Hamiltonian
\be
 H_{tr}=-\sum_{a,b}  \tau_{ab} c_a^\dagger c_b + \sum_{ab}\left[\Delta_{ab} c_a^\dagger c_b^\dagger + \text{H.C.} \right]
\ee
where the parameters are chosen so as to minimize the variational free energy
\be
F_{var}[\tau,\Delta] = F_{tr} + \langle H-H_{tr}  \rangle_{tr}
\ee
where the expectation values are taken with respect to the Gibbs ensemble of $H_{tr}$.  From the condition that $F_{var}$ is stationary with respect to variations of the parameters entering $H_{tr}$, we obtain the self-consistency conditions 
\be
\Delta_{ab} = \sum_{ a^\prime b^\prime}V_{ab b^\prime a^\prime}\langle c_{ b^\prime} c_{ a^\prime}\rangle
\label{self}
\ee
and a corresponding self-consistency conditions for $\tau_{ab}$.

Still at mean-field level, the superconducting state is characterized by a broken symmetry, and thus occurs whenever the lowest free energy solution of these equations has a non-vanishing value of $\Delta_{ab}$ for any $ab$;  the mean-field $T_c$ is  the upper bound on temperatures at which such a solution exists.   If the mean-field transition is continuous, $T_c$ can be identified by studying solutions of the linearized gap equation. 
Specifically, the pair field expectation value to linear order is given by
\be
\langle c_ac_b\rangle =\frac{\partial F_{tr}}{\partial \Delta^\star_{ab}}= - \chi_{ab, a^\prime b^\prime} (T)\ \Delta_{a^\prime b^\prime}\ +\  {\cal O}(|\Delta|^2\Delta^\star)
\ee
where the notation is such that repeated indices are summed over, and $\chi$ is the pair-field susceptibility computed in the normal state (\emph{i.e.} in the thermal ensemble of $H_{tr}$ with $\Delta_{ab}=0$). 
It is easy to see that $\chi$ is a Hermitian, positive semidefinite matrix 
 so the matrix square-root  $\chi^{1/2}$ 
exists and is also Hermitian.

With this in mind, we can expand the self-consistency equation (\ref{self}) to linear order.  Specifically,    consider the solutions of the eigenvalue equation
\be
\lambda_n(T)  F^{(n)}_{ab} = 
M_{ab, b^\prime a^\prime }(T) \  F^{(n)}_{b^\prime a^\prime}
\label{lambda}
\ee
where $M$ is the real Hermitian matrix 
\be
M_{ab,cd} \equiv -
\chi^{1/2}_{ab,a^\prime b^\prime}(T)\ V_{a^\prime b^\prime,c^\prime d^\prime}\ \chi^{1/2}_{ c^\prime  d^\prime, cd}(T)
\ee
Ordering the eigenstates $\lambda_n\geq \lambda_{n+1}$, the mean-field $T_c$ is then identified as the temperature at which $\lambda_{max}(T_c) \equiv\lambda_{0}(T_c)=1$.  If the mean-field transition is first order, the transition temperature must always be greater than that inferred in this way; thus, $\lambda_{max}(T)=1$ is a lower-bound estimate of the mean-field $T_c$.

\section{Qualitative aspects of the solution}   \label{sec:soln}

The linearized gap equation can be solved numerically on moderately large systems for any given configuration, as discussed below.  Analytic solutions are difficult, but a few general observations allow for qualitative statements under broad circumstances.  

We consider a short-ranged interaction $V_{a_1a_2,a_3a_4}$ which vanishes sufficiently rapidly so as to be negligible for $|\vec r_i-\vec r_j| \gg 1$ for any $i\neq j$ (with lattice constant = 1).  However, the physics of metals is reflected in the behavior of $\chi$, which becomes increasingly long range as $T\to 0$.  Specifically, at $T=0$, in either a clean metal or a diffusive metal, $\chi_{aa,a^\prime a^\prime}  \sim |\vec R|^{-d}$ for $|\vec R| \gg 1$; 
  the fact that $\sum_{\vec R} \chi_{(\vec 0\uparrow),(\vec 0\downarrow),(\vec R\downarrow),(\vec R\uparrow)}$ is logarithmically divergent is  a direct manifestation of the Cooper instability of a Fermi liquid -- and the fact (Anderson's theorem) that it persists even in the presence of  disorder.\cite{anderson_1959, AG_1959}

Were we studying quantum phase transitions at $T=0$, the long-range character of $\chi$ would make a statistical analysis of $M$ particularly subtle  (see Ref. \onlinecite{spivak_2008}).  However, as we are studying finite temperature transitions, there is always a finite length scale, $\ell(T)$, beyond which $\chi$ falls exponentially.  Specifically, in a metal, $\ell(T)$ is determined by a thermal coherence length, $L_T$, that diverges with a power of $T$ as $T\to 0$:  as $L_T = \hbar v_F/T$ in a clean metal and as $L_T = \sqrt{\hbar D/T}$ in a diffusive metal where $v_F$ and $D$ are, respectively, the Fermi velocity and the electron diffusion constant.  If the disorder is sufficiently strong to produce localization (which is to say any disorder in 2d), the same considerations apply so long as $L_T<\xi_{loc}$ where $\xi_{loc}$ is the localization length, but at low enough temperatures, $L_T$ saturates to $ \xi_{loc}$.  In other words, $\ell(T) \approx {\rm{min}}[L_T,\xi_{loc}]$.  Thus, $M_{a_1a_2,a_3a_4}$ is short-ranged in the sense that it vanishes exponentially when $|\vec r_i-\vec r_j| \gg L_T$ for any $i\neq j$.\footnote{Even at finite $T$,  the fact that $\ell(T) \gg 1$ is an essential feature of the theory of superconductivity.  In all cases of interest, the interaction strength in the appropriate pairing channel is always small compared to $E_F$.  Thus, even the largest single matrix elements in $M$ are small in magnitude -- it is only the fact the range of $M$ diverges in a non-integrable fashion that gives rise to the Cooper instability in this limit.} 

\subsubsection*{Lifshitz tails and a  ``theorem''}
The problem of determining the spectrum of $M$ is thus structurally similar to the problem of a non-interacting quantum particle moving in a random potential.  We can view $M$ as a temperature-dependent tight-binding model with moderate but finite range random hopping terms extracted from a complicated disorder ensemble; the higher the temperature, the more local is $M$.  The generic structure of the solutions is familiar, and therefore without further analysis we can infer various properties of the present problem on the basis of well-established results from the random potential problem.  We will adopt the terminology that anything that can be established on the basis of this precise analogy is ``proven,'' although (in common with ``Anderson's theorem'') no proof in the mathematician's sense of propositions and lemmas will be attempted.  

On this basis, we conclude that there is a $T$-dependent ``density of states,'' $\rho(\lambda;T)$, for the eigenvalues of $M$, and that this distribution is self averaging, $\rho(\lambda) = \overline {\rho(\lambda)}$.  Where $\rho$ is large, we expect solutions to be delocalized or at most weakly localized on exponentially long length scales.  But likewise, there are universal ``Lifshitz tails'' to the distribution \cite{lifshitz_1965} that extend to values of  $\lambda$ well in excess of the mean.  In the tails, the eigenstates are strongly localized and are associated with unusual rare regions in which the configuration is exceptionally conducive to superconductivity.  The structure of this tail in the density of states depends on details of the disorder.  In many circumstances its asymptotic form, in the regime in which $\rho(\lambda;T)$ is approaching zero, can be estimated by solving an optimization problem \cite{lifshitz_1965, halperin_1966, zittartz_1966}, which in turn can be related to a replica symmetry breaking instanton solution of an appropriately replicated effective field theory. \cite{cardy}  

In the present problem, as in the problem originally analyzed by Lifshitz \cite{lifshitz_1965}, we expect that the distribution is bounded, \emph{i.e.} that $\rho(\lambda;T) \neq 0$ only within a finite range $\lambda_{min}(T)< \lambda < \lambda_{max}(T)$.  Consider all possible ways the system could be organized within the allowed ensemble, including all possible concentrations and local (possibly highly ordered) arrangements of impurity atoms; $\lambda_{max}(T)$ is then associated with the optimal configuration.  
Of course, the concentration of regions of the sample that happen to well approximate this optimal configuration is extremely small.  The more precisely we require the optimal configuration, and the larger the region in question, the smaller the concentration of such regions; however, in the thermodynamic limit, for any given size and required precision, the probability of finding such a region is non-zero, and hence $\rho(\lambda) >0$  so long as $\lambda < \lambda_{max}$. 

 {\it Thus, the mean-field critical temperature is bounded below by $T_{c,max}$, which is defined implicitly as the solution of $\lambda_{max}(T_{c,max}) = 1$. In other words, the mean-field transition of a disordered system is determined by  the highest possible $T_c$ of any system that is allowed by the physics!}  This is generically larger than $T_c$ of the average Hamiltonian, $\overline H$.  This is the  new ``theorem.''

\subsubsection*{Significance of states in the tails}
The eigenstates with $\lambda$ close to $\lambda_{max}$ are localized in rare regions with a local configuration peculiarly conducive to superconductivity.  For temperatures slightly smaller than $T_{c,max}$, these regions can be treated as isolated superconducting ``puddles.''  The magnitude of the order parameter $\Delta^{(n)}$ on the puddle $n$ -- which depends on the character of the non-linear terms in Eq. \ref{self} and so cannot be directly computed from the solutions of the linearized equation we have studied here -- can nonetheless be estimated to depend on $\lambda_n$ as 
 \be
 \Delta^{(n)}(T) \sim \Delta_0 \sqrt{\lambda_n(T) -1}. 
 \ee
So long as the puddles are both rare and uncoupled in this range of temperatures,  their effect on the global properties of the system will generally be relatively weak.  However, the presence of an increasing concentration $n_{sc}$ of locally superconducting regions with decreasing temperature could be detected in various ways:  local spectroscopy will see the opening of what should look like a superconducting gap in any region in which $\lambda_n(T) >1$.  

It is only when $n_{sc}(T) \ell^d(T) $ ceases to be small that the Josephson coupling between such superconducting puddles will start to become significant, so that the more familiar signatures of superconductivity, such as a significant drop in the resistivity and substantial diamagnetism, will appear.  Global phase coherence -- \emph{i.e.} the true $T_c$ -- is a still lower temperature where the puddles effectively begin to overlap.  Even though the electronic structure is that of a statistically homogeneous metal for $T>T_{c,max}$ and a correspondingly homogeneous superconductor for $T_c \ll T$, it is effectively  an inhomogeneous mixture of superconducting puddles embedded in a normal metal matrix for $T_c \lesssim T < T_{c,max}$.  In weakly coupled superconductors, in which $\xi_0$ is large compared to the correlation length of the disorder, this regime is parametrically small (by a factor more or less equivalent to the usual Ginzburg parameter).  However, for superconductors with a short correlation length, this regime is not small, in which case significant deviations from mean-field behavior are to be expected, and $T_c$ is determined more by phase ordering than by (local) gap formation. 

\subsubsection*{Approximate analytic considerations}
To get a feeling for the nature of the states in the Lifshitz tail, we make a variational ansatz for a localized eigenstate,
\be
F_{a_1a_2} =  f_{\sigma_1,\sigma_2}(\vec r_1-\vec r_2) F\left(\frac {\vec r_1+\vec r_2}2\right)
\ee
with the normalization condition $\sum_{\vec r, \sigma, \sigma^\prime}|f_{\sigma,\sigma^\prime}(\vec r)|^2 = 1$.
This ansatz assumes that there is a preferred form of the pair-wave-function, with the only variational freedom associated with the local magnitude, $F(\vec R)$.  The simplified eigenvalue equation is then
\be \label{eq:simpEig}
\sum_{\vec r^{\prime}} M(\vec r,\vec r^{\ \prime}) F(\vec r^{\ \prime}) = \lambda F(\vec r)
\ee
Moreover, if we further assume that $F$ is a slowly varying function over the range of $M$ (\emph{i.e.} $\ell$) then we can treat $\vec r$ as a continuous variable and expand the eigenvalue equation in a gradient expansion as
\be
\big[ \gamma(\vec r;T) + (1/2)\nabla_\mu W_{\mu,\nu}(\vec r;T)\nabla_\nu + \cdots \big] F(\vec r) = \lambda F(\vec r).
\ee
Finally, again invoking the central limit theorem, we can approximate $\gamma(\vec r;T) \approx \overline{\gamma(T)} + \delta \gamma(\vec r)$ where $\delta\gamma(\vec r)$ is a local gaussian random variable with variance $\overline{\delta\gamma(\vec r)\delta\gamma(\vec r^{\ \prime})}=\delta(\vec r-\vec r^{\ \prime})\sigma^2$, and $W_{\mu,\nu}(\vec r;T) \approx \overline{W_{\mu,\nu}(\vec r;T) } = \delta_{\mu,\nu}(1/m)$.  We moreover expect $m>0$ reflecting the fact that the superconducting susceptibility generally favors spatially uniform pair-wave-functions over oscillatory ones, and on dimensional grounds we expect $m \sim \ell(T)^{-2}$.  

This leaves us precisely with the problem of Lifshitz tails for a particle moving in a random potential, and  asymptotic forms of $\rho(\lambda)$ have been derived in several classic references. \cite{cardy, yaida}  For $d<4$, the essential features of the results can be derived rather simply as follows:  assuming we are interested in localized solutions with $\lambda > \overline \gamma$, we look to find a region of size $R$ in which the average value of $\delta \gamma$ is sufficiently positive to produce the desired eigenvalue.  This requires that the constraint $\delta\gamma =(\lambda - \overline \gamma) -\beta m^{-1} R^{-2}$ be satisfied where the precise value of $\beta$ depends on the shape of the assumed puddle -- it has its smallest value if the puddle is spherical.  The probability of finding such a puddle $\sim \exp[-S]$ where $S=\beta^\prime R^d (\delta\gamma)^2/[2\sigma^2]$ and $\beta^\prime R^d$ is the volume of the puddle.  Minimizing $S$ with respect to $\delta\gamma$ subject to the above constraint, yields
\be
S = \alpha \sigma^{-2} \ell^d \ [\lambda - \overline \gamma(T)]^{(4-d)/2}
\ee
and $R = \alpha^\prime \ \ell [\lambda -\overline \gamma(T)]^{-1/2}$ where $\alpha$ and $\alpha^\prime $ are numbers of order 1 (that can be expressed in terms of $\beta$, $\beta^\prime$, $m_0$, and $d$.).   Finally, if we assume that $\overline \gamma(T)$ can be approximated as a linear function of $T$ as $\overline \gamma(T) \approx 1 - \gamma_0(T-T_0)/T_0$, then this yields a concentration of superconducting regions
\bea
\label{analytic}
n_{sc}(T)  \sim && \rho(\lambda=1,T) \\
&& \sim \exp\left\{- A \left(\frac{\xi_0}{a}\right)^d \left(\frac {T-T_0}{T_0}\right)^{(4-d)/2}\right\} \nonumber
\eea
where $T_0$ is the average $T_c$, $\xi_0$ is the corresponding ($T=0$) superconducting coherence length, $a$ is the correlation length of the disorder, and $A\sim \gamma_0/\sigma^2$ is a constant that depends on the strength of the disorder and other  features we have swept under the rug.

\section{Some numerically solved examples}   \label{sec:numeric}

To see how the general considerations outlined above play out explicitly, we have computed $\rho(\lambda)$ by numerical solution of Eq. \ref{lambda} on a square lattice with periodic boundary conditions and  sizes $L\times L$ with $L$ between 24 and 50.  Typically we average our results over 600 distinct disorder configurations.  For the ``uniform'' problem we take $\overline H$ to be a nearest-neighbor tight-binding model on a square lattice with hopping matrix $\overline t=1$.  The chemical potential is chosen such that $n = 0.8$ electrons per site.  To obtain smooth curves for $\rho(\lambda)$, a gaussian broadening of the levels has been introduced with $\delta\lambda = 0.01$.

We define different disorder ensembles as follows:  1)  To represent potential disorder, we add a random single-site energy, $t_{(\vec r,\sigma),(\vec r^\prime,\sigma^\prime)} = \epsilon(\vec r) \delta_{\sigma,\sigma^\prime}\delta_{\vec r,\vec r^\prime}$;  when we take $\epsilon$ from a square distribution of width $W$ we will refer to it as ``Born scattering'', while the binary distribution $\epsilon = 100 \bar t$ or $\epsilon=0$ with probabilities $n_{\text{imp}}$ and $1-n_{\text{imp}}$, respectively, will be referred to as ``unitary scatterers''.  2) To represent the pairing interaction, we assume an on-site $V_0(\vec r)$ and nearest-neighbor $V_1(\vec r)$ density-density interaction.  The case in which we take $V_0 < 0$ and $V_1 = 0$ we will refer to as ``s-wave pairing,'' while the case with  $V_0 > 0$ and $V_1 < 0$ will be called ``d-wave.'' We consider both uniform ($V_0$ and $V_1$ independent of $\vec r$) and inhomogeneous pairing interactions, where for instance to represent ``negative U centers'' we take $V_0= -  U$ with probability $x$ and $V_0=0$ with probability $1-x$.  We work in units of hopping $\overline{t} = 1$ and take values of $U = 3, 4$.  Since $\lambda$ scales with $U$, we scale the x-axis by $|U|$ in the figures below such that the distribution is independent of our chosen value.

\begin{figure}
\includegraphics[width=3.4in]{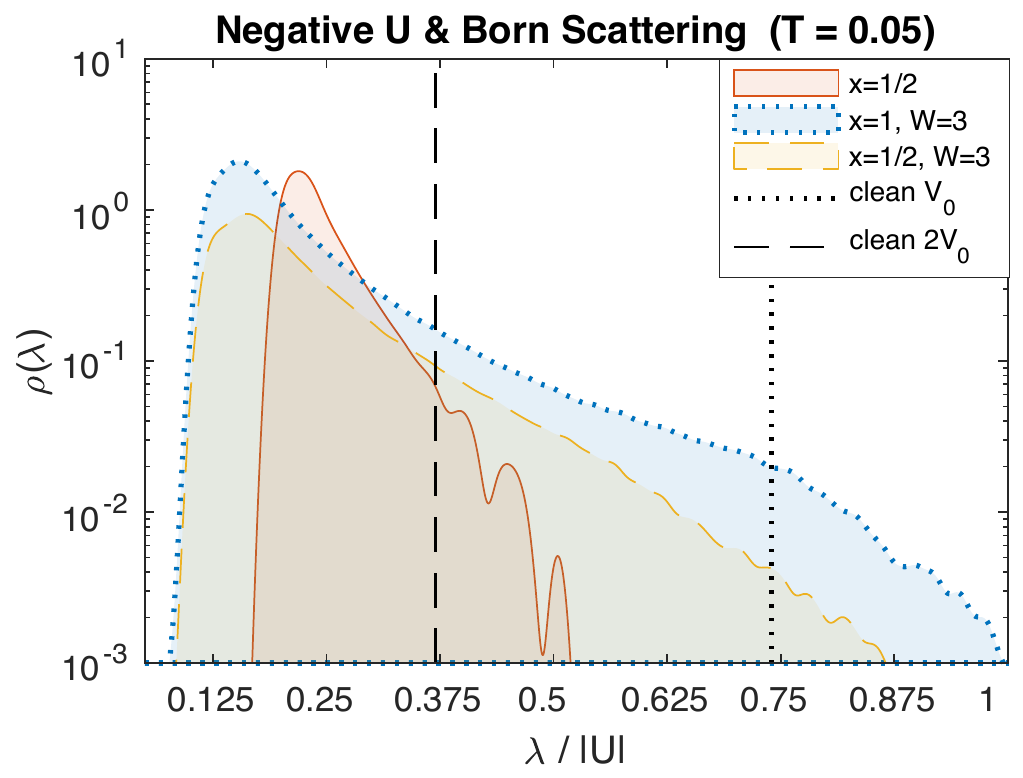}
\caption{Effect of disorder on ``s-wave'' pairing:  Log-linear plot of disorder-averaged $\rho(\lambda)$ with attractive on-site  (``s-wave'')  interactions ($V_1 = 0$) at temperature $T = 0.05$.  The blue-dashed curve is for uniform $V_0 = -U$ and Born scattering with $W = 3 \overline{t}$.  The other  curves are for the case of negative $U$ centers where $V_0=-U$ with probability $x=1/2$ and $V_0=0$ with probability $(1-x)$;  the yellow-dashed curve includes additional Born scattering with $W = 3 \overline{t}$ while the red-solid curve has no additional single-particle disorder ($W=0$).  The dotted vertical (black) line corresponds to the maximum eigenvalue of the disorder free system, $\lambda_{max}(W=0,V_0=-U,T=0.05)$, while the dashed vertical line represents $\lambda_{max}(W=0,V_0=-Ux,T=0.05)$ with $x=1/2$.}
\label{fig:S-wave_RHO}
\end{figure}

\subsubsection{Born scattering and s-wave pairing} 
The blue-dotted curve in Fig. \ref{fig:S-wave_RHO} shows $\rho(\lambda)$ for the case of uniform s-wave pairing ($V_0=-U$) and Born scattering with $W=3$ at temperature $T = 0.05$.  As we plot $\lambda$ in units of $U$, the results apply for arbitrary values of $U$.  For comparison, the dotted vertical lines represent the maximum value of $\lambda$ for the disorder free model ($W=0$):  $\lambda_{max}(W=0,T=0.05) / |U| = 0.74$.  Since the results are represented at fixed $T$, the meaning of the dotted curve is that for the disorder free system described by $\overline{H}$, the critical strength of $U$ such that $T_c > T$ for any $U > U_c(T)$ is determined by $U_c(T) = \overline{t} \ \lambda_{max}(W=0,T)$.  Manifestly, at $T=0.05$, the tail of the distribution in the presence of disorder is easily seen to extend well above the $\lambda_{max}$, meaning that disorder enhances $T_c$ as expected.  

 \begin{figure*}
\includegraphics[width=1.92 \columnwidth]{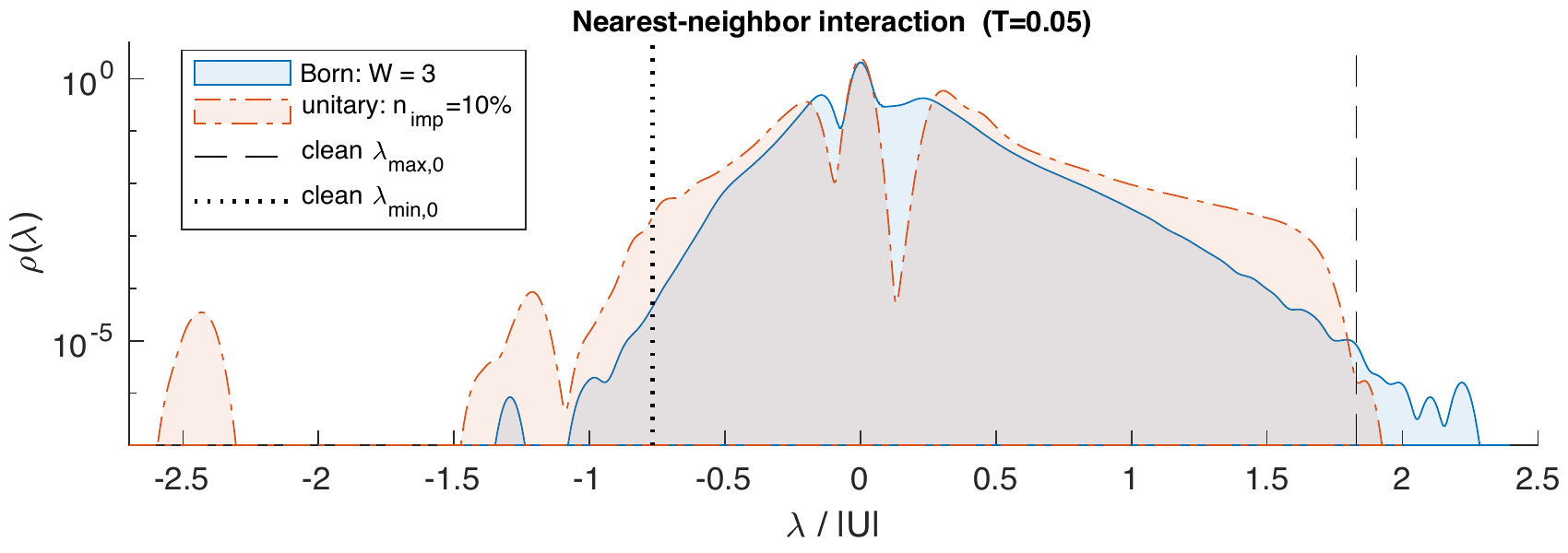}
\caption{Effect of disorder on ``d-wave'' pairing:  Disorder-averaged $\rho(\lambda)$ with uniform interactions, $V_0 = +U$, $V_1  = -1.5 U$ at $T = 0.05$.  The blue-solid line is for Born scattering with $W = 3$, and the red-dashed line is for unitary scattering with $n_{\text{imp}} = 10\%$.  The dashed and dotted vertical lines represent maximum and minimum eigenvalues, respectively for disorder free system, $W=0$ and $n_{\text{imp}} = 0$.}
\label{fig:d-RHO}
\end{figure*}

\subsubsection{Born scattering plus negative U centers} 
The red-solid and yellow-dashed curves in Fig. \ref{fig:S-wave_RHO} show $\rho(\lambda)$ for a concentration $x=1/2$ of negative $U$ centers in the absence and presence of Born scattering ($W=0$ and $W=3$), respectively.  Since the average strength of the attractive interaction is $\overline{V_0} = - U/2$, we have also, for comparison, shown as the vertical dashed line the maximum value of $\lambda$ one would obtain for a disorder free system with this average value of the effective interaction.  The integral of the distribution is normalized to unity, though half of the eigenvalues are exactly zero (not shown) when $x = 1/2$.  The tails of the distribution extend above the dashed line showing that disorder enhances $T_c$.

There is additional structure to the distribution with no Born scattering.  For example, we have checked that the separated peak at $\lambda / |U| \sim 0.5$ appears for system sizes between $24 \leq L \leq 50$ with over 1000 disorder configurations suggesting the peak is not due to insufficient statistics.  Such peaks in the distribution are associated with localized solutions associated with certain distinct sorts of local structures.  Similar peaks appear for $x=1/10$ where they can be associated with clusters of 6--8 negative U centers.  
 
\subsubsection{d-wave pairing with Unitary and Born disorder} 
Disorder-averaged $\rho(\lambda)$ at a temperature $T = 0.05$, electron density $n=0.8$, and uniform interactions of a sort that favor d-wave pairing, $V_0 = +U > 0$ and $V_1 = -1.5 V_0$, are shown in Fig. \ref{fig:d-RHO}.  Here the blue-solid line is for Born scattering with $W = 3$, while the red-dashed line is for a concentration $n_{\text{imp}} = 10\%$ of unitary scatters.  For comparison, the dashed and dotted vertical lines correspond, respectively, to the maximum and minimum eigenvalues of the disorder free reference system ($W=0$ and $n_{\text{imp}}=0$), $\lambda_{max}(W=0,n_{\text{imp}}=0,T=0.05) / |U|= 1.83$ and $\lambda_{\text{min}}(W=0,n_{\text{imp}}=0,T=0.05) / |U|= -0.77$.  

Notice that although $\rho(\lambda)$ is very small by the time $\lambda$ is comparable to $\lambda_{max}$, in both the Born and the unitary case the distribution extends beyond this point.  Thus, even in the case of ``d-wave'' pairing, the mean-field $T_c$ is enhanced by disorder, as promised.  As in the previous example, there is additional structure to the distribution superimposed on a generally rapid decay at large $\lambda$.  For instance, we have checked that the multiple peaks seen in the Born case at the largest values of $\lambda$ are not much affected by system size nor with the change in the number of disorder configurations over which the average is taken.  Here, the corresponding eigenstates are moderately well localized, and the peaks represent structures associated with particular classes of local disorder configurations. 

If, as assumed, $V_0>0$, the solutions with negative $\lambda$ are of no physical significance.  However, the same calculation applies to the case of $V_0<0$, where the model corresponds to an attractive on-site interaction with a weaker repulsive nearest-neighbor repulsion.  In this case, the role of positive and negative $\lambda$ are interchanged.  Of course, in this case, the pairing is s-wave, and not surprisingly the results are quite reminiscent of these for the negative $U$ problem.  Here, we can see an even more vivid example of local structures favoring highly localized large $\lambda$ solutions;  the eigenstates corresponding to the peak in the distribution for the unitary scatterers at $\lambda / |U| \sim -2.5$ are typically localized in a radius of only a few lattice sites.

\section{Relation to Other Work}   \label{sec:otherwork}



Needless to say, the effect of disorder on the superconducting transition temperature,  the character of superconducting fluctuations, and the structure of the superconducting state have been the subject of extensive theoretical investigation for many decades.  Since for the most part, these studies had in mind large coherence-length superconductors, the effects of inhomogeneous pairing were (rightly) largely neglected in these studies.  However, more recently, especially since the discovery of cuprate high temperature superconductivity, a number of studies have highlighted various circumstances in which inhomogeneous pairing correlations play a significant role in superconductors with short correlation lengths  -- see, for example, Refs. \onlinecite{podolsky_2005, yukalov_1995, yukalov_2004, garcia_2014, scalettar_2006, trivedi_1998, trivedi_2001, nunner_2005, andersen_2006}.   These effects are particularly amplified near a $T=0$ (quantum) superconductor to metal\cite{spivak_2008, larkin_1998} or superconductor to insulator\cite{trivedi_1998, trivedi_2001, trivedi_2011} transition.  Circumstances in which disorder enhances $T_c$ have likewise been discussed previously.\cite{garcia_2015, feigelman_2007, mirlin_2012, mirlin_2015, franz_1997, walker_1998, gastiasoro_2017, romer_2017}  For instance, in a semimetal with weak attractive interactions, $T_c$ is zero in the absence of disorder, but non-zero in the presence of disorder.\cite{sondhi_2013, sondhi_2014}  However, as far as we know, that universal Lifshitz tails give rise to a broad distribution of local mean-field $T_c$'s, and the implied importance of thermal (classical) phase fluctuations for the thermal transition in short-coherence length superconductors, has not been sharply articulated previously.

\section{Possible relevance to the cuprates}  \label{sec:cuprates}

Of course, the best known short-coherence length superconductors are the cuprates in which, depending somewhat on the range of doping and the method used to make the estimates, the superconducting coherence length is thought to vary from a few lattice constants up to perhaps a dozen lattice constants.  Moreover, the fact that these materials are highly quasi-2D, means that the large factor of $(\xi_0/a)^d$ in the exponent of Eq. \ref{analytic} has $d=2$ rather than $d=3$ which enhances the range of temperatures over which the present considerations are relevant.  With that in mind, we note that there are several features of the body of experimental observations that indeed may reflect a degree of self-organized granularity in the superconducting phenomena.

Inhomogeneity in the gap has been observed in scanning tunneling microscopy measurements of peaks in the local density of states in BSCCO.\cite{howald_2001, mcelroy_2005, yazdani_2007}  
 Significantly, it has been observed\cite{yazdani_2007} that regions that locally have large values of the gap at low $T$ have a local gap onset temperature that is larger than average, roughly in proportion to the local value of the low $T$ gap.  This is suggestive of the sort of local variations in the pair-wave-function discussed above.  Moreover, this correlates with the observation that $T_c$ is typically determined by the superfluid stiffness, rather than the pairing scale.\cite{emerykivelson, vishik_2012, zhong_2018}  On the other hand, the above mentioned results primarily concern the underdoped cuprates, where even the ``normal'' state above $T_c$ deviates dramatically from the usual Fermi liquid metal phase on which BCS theory is based.  Below the critical temperature, however, mean-field d-wave pairing inhomogeneity can still reproduce a relatively homogeneous LDOS at sufficiently low energy\cite{fang_2006} such that the proposed granularity above the critical temperature does not directly conflict with local spectroscopy in BSCCO observing a homogeneous superconducting state.\cite{seamus_2012, seamus_2015} 
 
Recently, interesting experiments have begun to systematically probe the overdoped regime, where the normal state is at least more nearly Fermi-liquid like.  Somewhat unexpectedly, it is found that the same relation between $T_c$ and the superfluid density persists, even as the quantum critical doping at which $T_c$ vanishes is approached.\cite{bozovic_2016}  Moreover, there is evidence both from optics\cite{armitage_2018} and specific heat\cite{hhwen_2004} that even deep in the superconducting state, a large density of apparently normal metallic quasiparticles persists, with a density that approaches that of the normal state as $T_c \to 0$.   We therefore consider it likely that these experiments reflect the sort of self-organized granularity that we have described here, though clearly local probe measurements are necessary to confirm this.  In making this suggestion, we wish to stress that what we are talking about is an intrinsic feature of short-coherence-length superconductors in the presence of statistically homogeneous disorder;  it has nothing to do with any large scale structural or chemical inhomogeneities that are typically meant when one discusses ``sample inhomogeneities.''
\footnote{A  superficially very different interpretation of these same experiments has been presented in Refs. \onlinecite{broun_2017} and \onlinecite{hirschfeld_2018}. Here, neglecting the role of inhomogeneous pairing correlations, the effective medium theory of disordered d-wave BCS superconductors has been deployed and shown to rather naturally account for many of the salient features  of the experiments below $T_c$  using plausible model parameters.  It is unclear to what extent the two approaches are truly at odds.  Both approaches start from the perspective that BCS mean-field theory is a plausible first approximation, and that the effects of disorder play an essential role even in the cleanest samples.  It is likely that   the induced self-organized inhomogeneity on which we have focused  is most significant above and in the vicinity of $T \approx T_c$;  below $T_c$, the familiar proximity effect tends to produce a more uniform electronic structure even in the presence of inhomogeneous pairing interactions.  However, close enough to the quantum critical point at which $T_c\to 0$, a variety of ``phase-sensitive'' effects could give rise to unambiguous macroscopic signatures of self-organized granularity, even in the superconducting state.\cite{spivakandme}}

\begin{acknowledgments}  We acknowledge useful discussions with D. Broun, J. C. Seamus Davis, P. Hirschfeld, and N. Trivedi.  This work was supported in part by the Department of Energy, Office of Basic Energy Sciences, under contract no. DE-AC02-76SF00515 at Stanford.
\end{acknowledgments}


\appendix 

\begin{center}
\noindent {\large {\bf Supplementary Material}}
\end{center}

\section{Linearized gap equation}

The spectrum of $M$ is computed numerically for different disorder realizations to extract the effect of rare events on mean-field $T_c$.  The disordered problem is diagonalized $H = \sum_{n} E_n \gamma^\dagger_n \gamma_n$ using $c_{r \sigma} = \sum_n U_{r \sigma ,  n }^* \gamma_n$ and Green's function
\begin{equation}
\langle c_{a \sigma}(\tau) c^\dagger_{b \lambda} \rangle = \sum_n  U^*_{a \sigma , n } U_{b \lambda , n } e^{ - \tau E_n } ( 1 - f_n )
\end{equation}
where the columns of $U_{r \sigma , n}$ are eigenvectors and $f_n = (1+e^{\beta E_n})^{-1}$.  For singlet operator defined as $\psi_{a b} = (c_{a \downarrow} c_{b \uparrow} + c_{b \downarrow} c_{a \uparrow} ) / 2$, the pair susceptibility is
\begin{widetext}
\[
\chi_{a b ; c d} = \int_0^\beta d \tau \ \big\langle \psi_{a b}(\tau) {\psi}^\dagger_{c d} \big\rangle  
= \frac{1}{2}  \int_0^\beta d \tau \  \big\langle  c_{a \downarrow}(\tau) c^\dagger_{c \downarrow}  \big\rangle \big\langle c_{b \uparrow}(\tau) c^\dagger_{d \uparrow} \big\rangle  + \big\langle  c_{a \uparrow}(\tau) c^\dagger_{d \uparrow}  \big\rangle \big\langle  c_{b \downarrow}(\tau) c^\dagger_{c \downarrow}  \big\rangle 
\]
\[
= \frac{1}{2} \sum_{ n , m = 1 }^{ 2 N_x N_y } \left( \frac{1 - f_n - f_m }{ E_n + E_m } \right) \Big[ U^*_{a \downarrow , n } U_{ c \downarrow , n } U^*_{b \uparrow , m } U_{ d \uparrow , m } + U^*_{a \uparrow , n } U_{d \uparrow , n } U^*_{b \downarrow , m } U_{ c \downarrow , m }   \Big] 
\]
\end{widetext}
for sites $a,b,c,d = 1, \cdots , N_x N_y$.  The interaction term has the form
\be
H_{\text{int}} = \sum_{r} V_0(r) c^\dagger_{r \uparrow} c^\dagger_{r \downarrow} c_{r \downarrow} c_{r \uparrow} + \sum_{\langle r , r' \rangle} V_1(r) \widehat{n}_r \widehat{n}_{r'}
\ee  
where $\widehat{n}_r = \sum_{\sigma} c^\dagger_{r \sigma} c_{r \sigma}$, and the inhomogeneous on-site interaction $V_0(r)$ and nearest neighbor interaction $V_1(r)$ are chosen to favor either s-wave ($V_0 < 0$ and $V_1 = 0$) or d-wave ($V_0 > 0$ and $V_1 < 0$) solutions.  The simplified eigenvalue equation (\ref{eq:simpEig}) is solved for $(5 N_x N_y) \times (5 N_x N_y)$ matrix $M_{r , r + \delta ; r' , r' + \delta' } =  V_{r,r+\delta}  \chi_{r , r+\delta ; r' , r' + \delta'}$ for $\delta, \delta' = \{ (0,0) , (\pm 1, \pm 1) \}$ and interactions $V_{r,r} = V_0(r)$ and $V_{r,r+\delta} = V_1(r)$.

\section{s-wave pairing at higher temperature}

\begin{figure}
\includegraphics[width=3.3in]{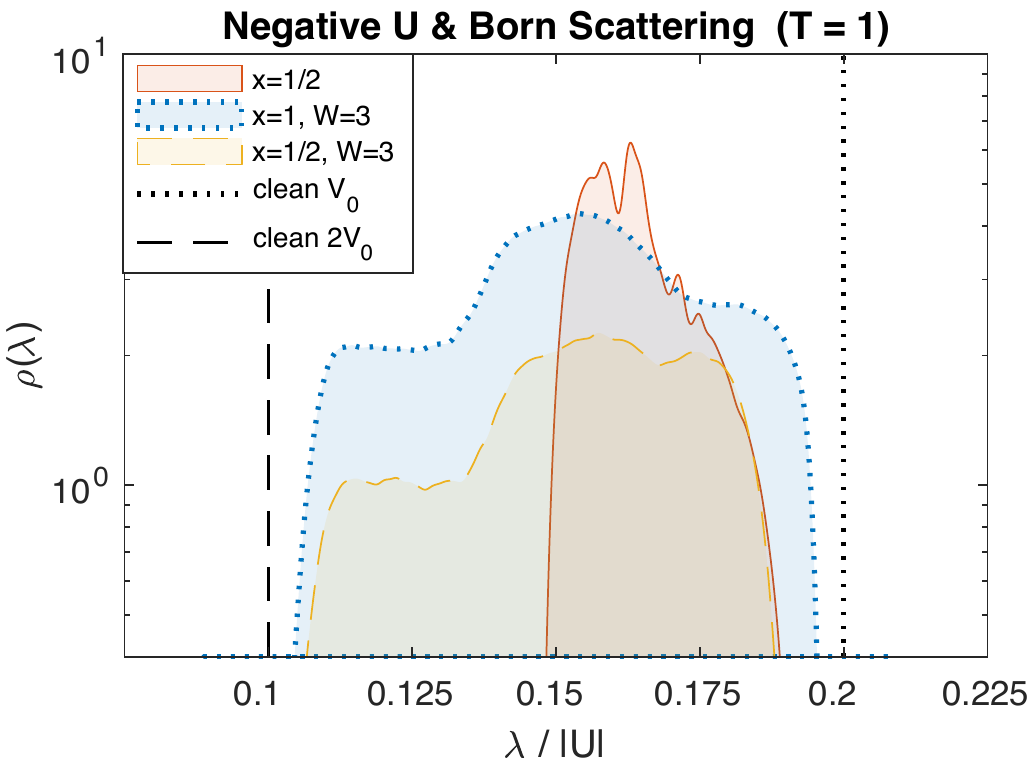}
\caption{Effect of disorder on ``s-wave'' pairing:  Log-linear plot of disorder-averaged $\rho(\lambda)$  
with attractive on-site  (``s-wave'')  interactions ($V_1 = 0$) at temperature $T = 1$.  The blue-dashed curve is for uniform $V_0=-U$ and Born scattering with $W=3\bar t$.  The other  curves are for the case of negative $U$ centers where $V_0=-U$ with probability $x=1/2$ and $V_0=0$ with probability $(1-x)$;  the yellow-dashed curve includes additional Born scattering with $W=3\bar t$ while the red-solid curve has no additional single-particle disorder ($W=0$).  The dotted vertical (black) line corresponds to the maximum eigenvalue of the disorder free system, $\lambda_{max}(W=0,V_0=-U,T)$, while the dashed vertical line represents $\lambda_{max}(W=0,V_0=-Ux,T)$ with $x=1/2$.}
\label{fig:S-wave_RHO_T-1}
\end{figure}

The disorder-averaged distribution $\rho(\lambda)$ at a higher temperature $T = \overline{t}$ is shown in Fig. \ref{fig:S-wave_RHO_T-1} for the same disorder ensembles shown in Fig. \ref{fig:S-wave_RHO}.  The maximum value of $\lambda$ for the disorder free model ($W = 0$) is $\lambda_{max}(W=0,T=1)/|U| = 0.2$.  For the high temperature case, the range of $M$ has dropped to being on the order of one lattice site, the fall-off of distribution is sufficiently sharp that the probability has dropped below the limits accessible to our numerical results before the clean-limit value is reached.  The increasingly fast drop of the tails of the distribution with increasing temperature is to be expected from the above considerations;  our failure to see the tail of the distribution at high temperatures extending above $\lambda_{max}$ is, we believe, an artifact but one that could only be corrected by keeping a much larger number of disorder configurations.  
For the case with negative U centers ($x=1/2$), the entire observed distribution lies above the dashed line. 

\section{Inverse Participation Ratio}

The degree to which the gap solution is localized can be quantified using the inverse participation ratio $I_\lambda$ for a given eigenvalue $\lambda$ defined as
\be
I_\lambda = \frac{\langle |F_\lambda(\vec{r})|^4 \rangle}{\langle |F_\lambda(\vec{r})|^2 \rangle^2}
\ee
where $\langle \cdots \rangle$ denotes a spatial average and $F_\lambda(\vec{r})$ is the eigenvector of the linearized BCS equation.  The inverse participation ratio has the property that $I_\lambda \sim 1 / L^d \rightarrow 0$ as linear system size $L \rightarrow \infty$ for extended states, and $I_\lambda \sim 1/\xi_{loc}^d$ for states localized within a characteristic length $\xi_{loc}$ as $L \rightarrow \infty$.  The inverse participation ratio therefore provides a length scale $1/\sqrt{I_\lambda}$ which gives an estimate $n_{sc} / I_\lambda$ of the degree to which superconducting solutions are isolated from each other in two dimensions.

{\bf Negative U centers:} The inverse participation ratio $I_\lambda$ decreases with increasing $\lambda$ meaning the solutions with the largest eigenvalues are most delocalized.  The characteristic length scale $1/\sqrt{I_\lambda}$ increases smoothly to roughly $L/2$ at a value $\lambda \sim 1.8$ at which point there is a jump in both $\lambda$ and $1/\sqrt{I_\lambda}$ (visible as a peak at $\lambda / V_0 \approx 0.5$).  

\begin{figure}
\includegraphics[width=3.5in]{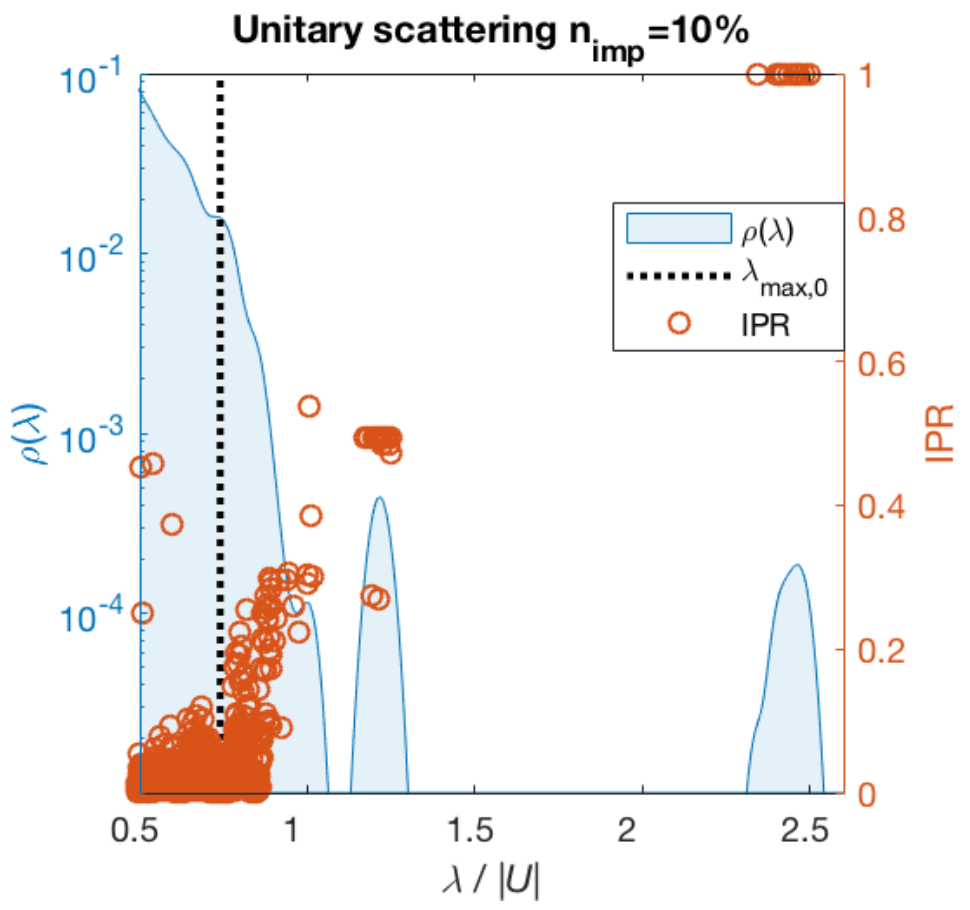}
\caption{Disorder-averaged $\rho(\lambda)$ (blue) for $x=1$, $V_0 = -4$, and $V_1 = 0$ with $n_{\text{imp}} = 10\%$ unitary scattering at $T = 0.05$.  The maximum clean eigenvalue $\lambda_{max,0}$ is represented by vertical dashed line (black), and inverse participation ratio by circles (red). Solutions with largest $\lambda / V_0 \sim 2.3$ have maximum inverse participation ratio $I_\lambda = 1$ and correspond to impurity band of states localized on a single site.  A second impurity band appears around $\lambda / V_0 \sim 1.25$ with $I_\lambda = 1/2$ corresponding to a solution localized on two sites.}
\label{fig:d-RHO_IPR}
\end{figure}

{\bf Unitary scattering with on-site interaction:}  Different types of disorder are more conducive to local enhancements of the density of states and therefore larger eigenvalues of $M$.  The results for $V_0 = -4$ and $V_1 = 0$ in the presence of pairing inhomogeneity $x = 1/2$ and a concentration $n_{\text{imp}} = 10\%$ of randomly located unitary scatterers of strength $V_{\text{imp}} = 100 t$ are shown in Fig. \ref{fig:d-RHO_IPR} at a temperature $T = 0.05 t$.  The inverse participation ratio for each of the eigenvectors is plotted (red circles) on top of the tail end of the $\rho(\lambda)$ distribution (blue).  The vertical dashed line (black) corresponds to the maximum eigenvalue $\lambda_{max,0}$ for the clean system at $T = 0.05 t$.  Above $\lambda \approx 1$ in the exponential tail of the distribution, the inverse participation ratio remains small corresponding to relatively extended states; however, above clean system maximum $\lambda_{max,0} \approx 3$, the solutions with the largest $\lambda$ show an increase in IPR.  The maximum eigenvalue solutions with $\lambda / V_0 \sim 2.3$ have $I_\lambda = 1$ meaning the gap eigenvector is localized on a single site.  From visual inspection of the solution and local density of states (see Supplementary Material, "LDOS Enhancement"), it can be seen that these solutions are associated with clusters of multiple unitary scatterers in close proximity.  There appear to be two peaks near $\lambda \approx 5$ and $\lambda \approx 9$ that can correspond to impurity bands with IPR = 1/2 (states localized on two sites) and IPR = 1 (states localized on single site) respectively, which would hybridize at sufficiently large concentration.  For $\lambda < 1$ (not shown), the IPR spans a range from zero to one; however, since the density of states $\rho(\lambda)$ is large, the IPR loses meaning due to mixing in the self-consistent mean-field solution.  The distribution at higher temperature $T = 1$ (not shown) similarly exhibits two peaks above the clean $\lambda_{max,0}$.

\section{LDOS Enhancement}

\begin{figure}
\includegraphics[width=3.5in]{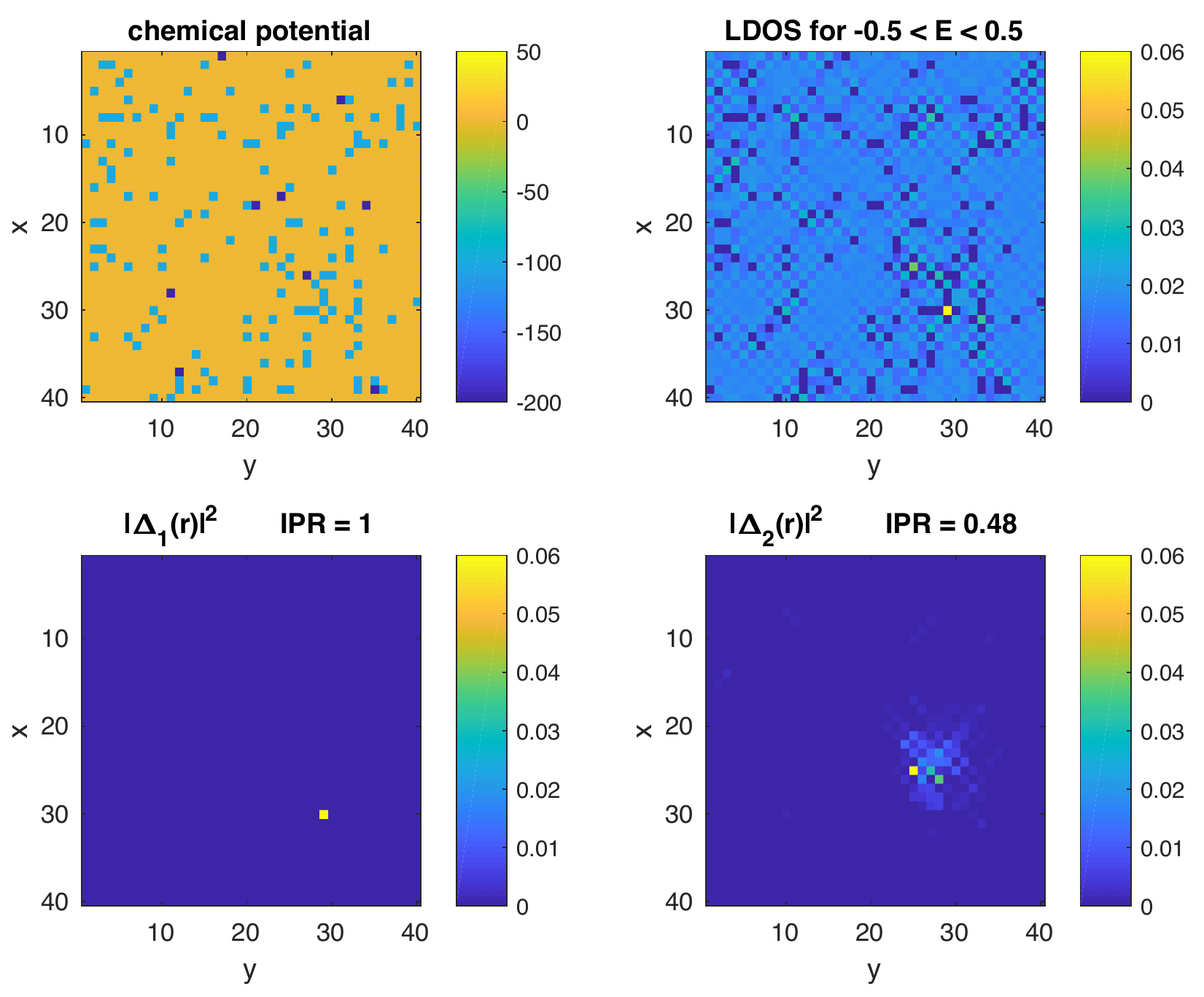}
\caption{Chemical potential, LDOS, and on-site pairing solutions with $n_{\text{imp}} = 10\%$ unitary scatterers for largest ($\lambda_1 = 9.7$) and second largest ($\lambda_2 = 3.4$) eigenvalues for a particular disorder realization at $T = 0.05 t$.  The peak in the local density of states is associated with a rare impurity cluster.}
\label{fig:ldos}
\end{figure}

From visual inspection of the gap solution and local density of states, it can be seen that there is an enhancement due to rare microscopic impurity configurations.  The chemical potential, local density of states, and on-site pairing solutions $F_\lambda(\vec{r})$ with $n_{\text{imp}} = 10\%$ unitary scatterers for largest ($\lambda_1 = 9.7$) and second largest ($\lambda_2 = 3.4$) eigenvalues for a particular disorder realization at temperature $T = 0.05 t$ are shown in Fig. \ref{fig:ldos}.  The on-site interaction is taken as $V_0 = -4$ and nearest neighbor interaction $V_1 = 0$.  The peak in the local density of states is associated with a rare impurity cluster, and the largest eigenvalue solution $F_{\lambda_1}(r)$ has $I_{\lambda_1} = 1$ corresponding to non-zero value on a single site.  The second largest eigenvalue solution $F_{\lambda_2}(\vec{r})$ for this disorder realization has $I_{\lambda_2} = 0.48$ and can be seen to have finite value over a localized region.

\

\

\section{Potential disorder with \\
nearest neighbor interaction}

The results for random potential and interaction disorder with nearest-neighbor interaction show qualitatively similar behavior to the case $V_1 = 0$:  the distributions decay exponentially with the eigenvalues in the tail exceeding the uniform clean case, and increasing temperature narrows the distribution.  In contrast to the previous example, the negative U centers with $V_1 \neq 0$ at high temperature $T = 1$ show two separated peaks:  the peak with larger eigenvalues corresponds to solutions localized in clusters where the interaction is non-zero such that the effective interaction of the puddle is $U$.  The peak with smaller eigenvalues (absent for $V_1 = 0$) corresponds to solutions localized on ``checkerboard'' patterns where the interaction averages to $U/2$ - the eigenvalues associated with this second peak are roughly half that of the larger-eigenvalue peak (though the lower end of this second peak has a range of effective interactions less than $U$).  This characterization is confirmed quantitatively by defining an effective gap $\overline{V}_1 = \sum_r V_1(r) |\Delta_x(r)|^2 / \sum_{r'} |\Delta_x(r')|^2$ of the puddle which shows $\overline{V_1} \approx V_1 / 2$ for states in the smaller-eigenvalue peak.   



\begin{thebibliography}{99}  

\bibitem{nunner_2005}
T. S. Nunner \emph{et al.} \prl \textbf{95}, 177003 (2005).

\bibitem{andersen_2006}
B. M. Andersen \emph{et al.} \prb \textbf{74}, 060501 (2006).

\bibitem{anderson_1959}
P. W. Anderson, J. Phys. Chem. Solids \textbf{11}, 26 (1959).

\bibitem{AG_1959}
A. A. Abrikosov and L. P. Gor'kov, JETP \textbf{9}, 220 (1959).

\bibitem{spivak_2008}
B. Spivak, P. Oreto, and S. A. Kivelson, \prb \textbf{77}, 214523 (2008).


\bibitem{lifshitz_1965}
I. M. Lifshitz, Sov. Phys. Usp. \textbf{7}, 549 (1965).

\bibitem{halperin_1966}
B. I. Halperin and M. Lax, Phys. Rev. \textbf{148}, 722 (1966).

\bibitem{zittartz_1966}
J. Zittartz and J. S. Langer, Phys. Rev. \textbf{148}, 742 (1966).

\bibitem{cardy}
J. L. Cardy, J. Phys. C \textbf{11}, L321 (1978).

\bibitem{yaida}
S. Yaida, \prb \textbf{93}, 075120 (2016).


\bibitem{podolsky_2005}
I. Martin, D. Podolsky, and S. A. Kivelson, \prb \textbf{72}, 060502 (2005).

\bibitem{yukalov_1995}
A. J. Coleman, E. P. Yukalova, V. I. Yukalov, Physica C \textbf{243}, 76 (1995).

\bibitem{yukalov_2004}
V. I. Yukalov and E. P. Yukalova, \prb \textbf{70}, 224516 (2004).

\bibitem{garcia_2014}
J. Mayoh and A. M. Garcia-Garcia, \prb \textbf{90}, 134513 (2014).

\bibitem{scalettar_2006}
K. Aryanpour \emph{et al.} \prb \textbf{73}, 104518 (2006).

\bibitem{trivedi_1998}
A. Ghosal, M. Randeria, and N. Trivedi, \prl \textbf{81}, 3940 (1998).

\bibitem{trivedi_2001}
A. Ghosal, M. Randeria, and N. Trivedi, \prb \textbf{65}, 014501 (2001).

\bibitem{larkin_1998}
M. V. Feigelman and A. I. Larkin, Chem. Phys. \textbf{235}, 107 (1998).

\bibitem{trivedi_2011}
K. Bouadim \emph{et al.}, Nature Physics \textbf{7}, 884 (2011).

\bibitem{garcia_2015}
J. Mayoh and A. M. Garcia-Garcia, \prb \textbf{92}, 174526 (2015).

\bibitem{feigelman_2007}
M. V. Feigel'man \emph{et al.} \prl \textbf{98}, 027001 (2007).

\bibitem{mirlin_2012} 
I. S. Burmistrov, I. V. Gornyi, and A. D. Mirlin, \prl \textbf{108}, 017002 (2012)

\bibitem{mirlin_2015}
I. S. Burmistrov, I. V. Gornyi, and A. D. Mirlin, \prb \textbf{92}, 014506 (2015).

\bibitem{franz_1997}
M. Franz \emph{et al.} \prb \textbf{56}, 7882 (1997).

\bibitem{walker_1998}
M. E. Zhitomirsky and M. B. Walker, \prl \textbf{80}, 5413 (1998).

\bibitem{gastiasoro_2017}
M. N. Gastiasoro and B. M. Andersen, ``Enhancing Superconductivity by Disorder,'' arXiv:1712.02656 (2017).

\bibitem{romer_2017}
A. T. Romer, P. J. Hirschfeld, and B. M. Andersen, ``Boosting $T_c$ with Disorder in Spin-Fluctuation Mediated Unconventional Superconductors,'' arXiv:1712.07914 (2017).



\bibitem{sondhi_2013}
R. Nandkishore \emph{et al.} \prb \textbf{87}, 174511 (2013).

\bibitem{sondhi_2014}
I. D. Potirniche \emph{et al.} \prb \textbf{90}, 094516 (2014).


\bibitem{howald_2001}
C. Howald, P. Fournier, and A. Kapitulnik, \prb \textbf{64}, 100504 (2001).

\bibitem{mcelroy_2005}  
K. McElroy \emph{et al.} Science \textbf{309}, 1048 (2005).

\bibitem{yazdani_2007}
K. K. Gomes \emph{et al.} Nature \textbf{447}, 569 (2007).

\bibitem{emerykivelson}
V. J. Emery and S. A. Kivelson, Nature \textbf{374}, 434 (1995).

\bibitem{vishik_2012}
I. M. Vishik \emph{et al.}, Proc. Natl. Acad. Sci. \textbf{109}, 18332 (2012).

\bibitem{zhong_2018}
Y. G. Zhong \emph{et al.}, ``Continuous doping of a cuprate surface: new insights from in-situ ARPES,'' arXiv:1805.06450 (2018).

\bibitem{fang_2006}
A. C. Fang \emph{et al.} \prl \textbf{96}, 017007 (2006).

\bibitem{seamus_2012}
K. Fujita \emph{et al.} J. Phys. Soc. Jap. \textbf{81}, 011005 (2012).

\bibitem{seamus_2015}
K. Fujita \emph{et al.} (2015) ``Spectroscopic Imaging STM: Atomic-Scale Visualization of Electronic Structure and Symmetry in Underdoped Cuprates,'' In: Avella A., Mancini F. (eds) \emph{Strongly Correlated Systems}. Springer Series in Solid-State Sciences, \textbf{180}. Springer, Berlin, Heidelberg




\bibitem{bozovic_2016}
I. Bozovic \emph{et al.} Nature \textbf{536}, 309 (2016).

\bibitem{armitage_2018}
F. Mahmood, X. He, I. Bozovic, and N. P. Armitage, ``Locating the missing superconducting electrons in overdoped cuprates,'' arXiv:1802.02101 (2018).

\bibitem{hhwen_2004}
H. H. Wen \emph{et al.} \prb \textbf{70}, 214505 (2004).

\bibitem{broun_2017}
N. R. Lee-Hone, J. S. Dodge, and D. M. Broun, \prb \textbf{96}, 024501 (2017);  Erratum-ibid. \textbf{97}, 219903 (2018).

\bibitem{hirschfeld_2018}
N. R. Lee-Hone \emph{et al.} ``Optical conductivity of overdoped cuprate superconductors: application to LSCO,'' arXiv:1802.10198 (2018).








\bibitem{spivakandme}  S. A. Kivelson and B. Spivak, \prb {\bf 92}, 184502 (2015).

\end{thebibliography}
 \end{document}